\documentclass[aps,twocolumn,preprintnumbers,amsmath,amssymb,prd,floatfix,nofootinbib,superscriptaddress]{revtex4-2}

\usepackage{amsmath,amssymb}
\usepackage[usenames,dvipsnames]{color}
\usepackage{graphicx}
\usepackage{subfigure}
\usepackage{soul}
\usepackage{hyperref}
\usepackage{multirow}
\hypersetup{colorlinks, linkcolor = [rgb]{0,0.0,0.75}, citecolor = [rgb]{0,0.0,0.75}, urlcolor = [rgb]{0,0.0,0.75}}

\newcommand\bef{\begin{figure}}
\newcommand\eef[1]{\label{fg:#1}\end{figure}}
\newcommand\beq{\begin{equation}}
\newcommand\eeq[1]{\label{#1}\end{equation}}
\newcommand\beqa{\begin{eqnarray}}
\newcommand\eeqa[1]{\label{#1}\end{eqnarray}}
\newcommand\bet{\begin{table}}
\newcommand\eet[1]{\label{tb:#1}\end{table}}

\begin{document}
\title{Semi-supervised learning of order parameter in 2D Ising and XY models using Conditional Variational Autoencoders}

\author{Adwait Naravane}
\email{abn19ms151@iiserkol.ac.in}
\affiliation{Department of Physical Sciences, \\ Indian Institute of Science Education and Research, Kolkata, India.}

\author{Nilmani Mathur}
\email{nilmani@theory.tifr.res.in}
\affiliation{Department of Theoretical Physics, \\
Tata Institute of Fundamental Research, 1 Homi Bhabha Road, Mumbai 400005, India.}

\preprint{TIFR/TH/23-13}

\begin{abstract}
We investigate the application of deep learning techniques employing
the conditional variational autoencoders for semi-supervised learning of
latent parameters to describe phase transition in the two-dimensional (2D)
ferromagnetic Ising model and the two-dimensional XY model. For both
models, we utilize spin configurations generated using the Wolff
algorithms below and above the critical temperatures.  For the 2D
Ising model we find the latent parameter of conditional variational
autoencoders is correlated to the known order parameter of
magnetization more efficiently than their correspondence in
variational autoencoders used previously. It can also clearly identify
the restoration of the $\mathbb{Z}_2$ symmetry beyond the critical
point.  The critical temperature extracted from the latent parameter
at larger lattices are found to be approaching its correct
value. Similarly, for the 2D XY model, we find our chosen network with
the latent representation of conditional variational autoencoders is
equally capable of separating the two phases between the high and low
temperatures, again at the correct critical temperature with
reasonable accuracy. Together these results show that the latent
representation of conditional variational autoencoders can be employed
efficiently to identify the phases of condensed matter systems,
without their prior knowledge.

\vspace*{0.3in}

\end{abstract}
\maketitle

\section{Introduction}
Over the past decade, the field of Artificial Intelligence (AI) has witnessed remarkable advancements driven by powerful computers and artificial neural networks. Artificial neural networks possess the ability to approximate virtually any existing function, making them versatile tools \cite{cybenko1989approximation}. These networks have already demonstrated their efficacy in diverse domains, such as image and speech recognition. Naturally, there is a growing interest in leveraging deep learning models, including generative neural networks, to tackle various physics problems. A certain class of neural networks called Physics-informed neural networks (PINNs) have been used to enhance information content by embedding the physical laws that govern the input data \cite{raissi2019physics}. One particular area of interest is the application of deep learning to condensed matter systems, especially those that can be formulated on a lattice and studied statistically. For example, restricted Boltzmann machines (RBMs), which are the simplest form of generative networks, have been employed to study phase structures and extract interesting features for various physical systems \cite{cossu2019machine}. Furthermore, several machine learning methods, such as supervised machine learning, RBMs, Autoencoders (AE) \cite{kramer1991nonlinear},  and Variational Autoencoders (VAE) \cite{kingma2013auto} have been able to successfully identify phase transitions in both for classical statistical mechanical systems \cite{carrasquilla2017machine}, such as the 2D Ferromagnetic and Anti-Ferromagnetic Ising models \cite{alexandrou2020critical} \cite{wetzel2017unsupervised}, and also for Quantum models such as the Hubbard model \cite{broecker2017machine}. Similar studies have also been conducted on lattice field theories such as the SU(2) gauge theory \cite{wetzel2017machine}. RBMs and deep Boltzmann machines have also been used for generating new Ising configuration \cite{morningstar2017deep}. Additionally, neural networks hold promise in accelerating traditional Monte Carlo algorithms near critical points \cite{wang2020neural}.

On the other hand, it is well-known that macroscopic properties of a system can be studied through microscopic investigation. In that, transitions between different phases
play an important role and the changes in order parameter can identify such phase transitions. It is thus an interesting question how machine learning can be utilized effectively to find such order parameters that are capable of detecting any transitions of macroscopic phases of a system. To achieve that, autoencoders (AEs) can be employed which are a type of neural network that can learn to represent complex high-dimensional data in a lower-dimensional latent space. They consist of an encoder that maps input data to a latent space and a decoder that maps latent vectors back to the original data space. Variational autoencoders, which have been developed recently by Diederik Kingma and Max Welling \cite{kingma2013auto}, are one of the best unsupervised machine learning algorithms. Unlike autoencoders which can learn a function to map the input data to a latent vector and a decoder as a reverse map, VAEs can learn a probability distribution of the data. VAEs are the generative extension of AEs \cite{blei2017variational}. VAEs have been applied in various fields, including speech recognition \cite{8268911}, computer vision \cite{bao2017cvae}, and natural language processing.  
 Utility of generative models, including VAEs, in generating physically meaningful configurations for Ising and XY models 
 have also been demonstrated \cite{d2020learning, cristoforetti2017towards}.

In our study, instead of using just VAEs, we employ unsupervised learning to explore the potential of a proposed fully connected (Dense) Conditional Variational  (C-VAE) \cite{sohn2015learning} in discovering a low-dimensional latent representation of the order parameter governing phase transitions in macroscopic systems. We choose the $2D$ Ising and XY models 
to demonstrate the efficacy of the method and our 
results show that the latent representation of C-VAE can be employed
efficiently to identify the phases of these systems, without their prior knowledge. For both models, we find the latent parameter of conditional variational
autoencoders is correlated to the known order parameter more efficiently than
their correspondence in variational autoencoders used previously.
CVAEs possibly can also be employed for other problems to identify their phases.

The paper is organized as below. In Sec. II, we discuss the Ising and XY models that we study. In Sec. III details of our methods using variational autoencoders are discussed. We present our findings in Sec. IV, and in Sec. V we provide our conclusion from this study and possible future outlooks.

\section{Models}
In this section, we first discuss the two models which are central to our work: $2D$ Ising model and $2D$ XY model. Of course these models are very well studied, but for the sake of completeness and for connecting to our methods we describe them briefly.  Next, we give the details of our training methods.
\subsection{2D Ising Model}
The Ising model is one of the most well-understood models in statistical mechanics. The Ising model in dimensions greater than one exhibits the ordered to unordered phase transition. The Hamiltonian of the Ising model \cite{peierls1936ising} in the absence of an external magnetic field is 
\begin{equation}
    H({s_i}) = -J \sum_{\langle ij \rangle} s_i s_j .
\end{equation}
Here, $S = \{s_i \}$ are the spin configurations and the notation $\langle ij \rangle$ denotes the summation over nearest neighbor.  The coupling term $J = 1$ and $-1$ are for the ferromagnetic and anti-ferromagnetic cases, respectively. The spin configuration for the 2D model can be represented by a matrix of form \cite{alexandrou2020critical},
\begin{equation}
    S = \begin{pmatrix} 
    \uparrow & \uparrow & \downarrow & \ldots & \uparrow \\ 
    \downarrow & \uparrow & \downarrow & \ldots & \uparrow \\ 
    \vdots & \vdots & \vdots & \ddots & \vdots \\ 
    \uparrow & \downarrow & \downarrow & \ldots & \downarrow  
    \end{pmatrix} .
\end{equation}
Thus a spin configuration can be taken as an image \cite{wetzel2017unsupervised} and hence is suitable for machine-learning.

The Ising model possesses the discrete $\mathbb{Z}_2$ symmetry which gets spontaneously broken below the critical temperature, and that transition can be studied through an order parameter. The magnetization of a spin configuration for a total number of spins  $N$ is given by,
\begin{equation}
    M({s_i}) = \frac{1}{N} \sum_i s_i .
\end{equation}
Using the expectation value of the above defined magnetization one can get a relevant order parameter at a fixed temperature as below,
\begin{equation}
    \langle M(T) \rangle = \frac{1}{Z} \sum_{{s_i}} M({s_i}) e^{-H({s_i})/T},
\end{equation}
where $Z$ is the partition function of the system. One generally uses the magnetization as the order parameter to identify the phase transition in this model. In this work, we use square lattices with periodic boundary condition and lengths up to $L = 70$. We use the Wolff cluster algorithm to generate 40200 samples for each of the 200 temperatures in the range $T \in [1, 4.5]$. 
\subsection{2D XY model}
The Hamiltonian of the XY model is defined as, 
\begin{equation}
    H({s_i}) = -J \sum_{\langle ij \rangle} \vec{s}_i . \vec{s}_j ,
\end{equation}
where the spins $s_i \in \mathbb{R}^2$. Interestingly for this model there is no continuous phase transitions in dimensions $d \leq 2$ which can be understood using  the Mermin-Wagner theorem. In 3D, XY model has 
a second order phase transition and has been studied thoroughly. In 2D however, we do not observe a second order phase transition associated with any symmetry breaking. Instead, one observes a high temperature disordered phase and a low temperature quasi-ordered phase with power law correlation below a certain critical temperature \cite{kosterlitz1973ordering}. 

In this study we use square lattices and sample 200 configurations each for 30 temperature points in the range $T \in [0, 2]$ using Wolff algorithm. The order parameter here is similar to that of Ising model with an $L^2$-norm consisting of two components of the magnetization vector.

\section{Methods}
In this section, we describe the neural networks that we utilized and its implementation for the Ising and XY models. A detailed discussion on these networks are given in the Appendix.

Autoencoders are a variant of traditional feed-forward neural networks used for
learning data in an unsupervised manner \cite{autoenc}. Conditional autoencoders are
modified autoencoders such that the input of the decoder is the output of the encoder along with a label for the input data \cite{sohn2015learning}. Variational autoencoders (VAE) are a version of autoencoders that impose additional constraints on the en-
coded representation \cite{doersch2016tutorial, kingma2019introduction}.  VAEs learn the parameters of a probability distribution modeling the data. This helps to generate samples closely resembling the input data once the network is trained. The main idea behind VAEs is to learn
a probability distribution over the latent space that can
generate data samples similar to the input data. This is
achieved by minimizing the {\it reconstruction error} between
the original data and the reconstructed data, while also
minimizing the \textit{Kullback-Liebler} (KL) divergence \cite{KL-d} between the learned latent distribution and a prior distribution (see the Appendix for more details).

\subsection*{Implementation of C-VAEs}
For the Ising model, we use a variational autoencoder with an encoder with three fully connected hidden layers of 625, 256, and 64 neurons, respectively \cite{alexandrou2020critical}. For the decoder, we also choose three hidden layers with 64, 256, and 625 neurons. We use a single order parameter for the Ising model. The activation function for the hidden layers is rectified linear units,  and we choose a linear activation function for the latent layer. We choose the activation function for the final layer of the decoder to be sigmoid in order to predict the probabilities of spins. 

For the XY model, we include a fully connected hidden layer with 256 neurons for both the encoder and decoder networks. In this case the number of latent parameters is two and the activation for the hidden layer and that of the latent layer is chosen accordingly. The activation function for the final layer of the decoder is chosen to be $tanh$ for predicting continuous values of the spin components. 

For the conditional VAE that we employ, we have taken the temperature of the configuration $T$ as the label $y$ associated with each input spin configuration. No dropout regularization scheme is employed. Instead, we use the KL divergence as regularization for the VAE. The reconstruction loss is the binary cross entropy for the Ising model and the mean-squared error for the XY model.  The size of the input and the output layers are $L \times L$. For the Ising model, the input consists of binary numbers corresponding to spin down and spin up. For the XY model, the input and output layers are of size $2 \times L \times L$ with normalized X and Y components of individual spin.

For training of the network, we split the data into training $(70 \%)$ and testing $(30 \%)$ sets. We then performed the training for between 50 to 300 epochs depending on the input size with varying batch sizes to maximize accuracy. We find that 50 epochs and a batch size of 64 give the best latent parameter for the 2D XY model. We optimize the learning rate of the network using the Adam optimizer \cite{kingma2014adam}. The implementation of the network has been performed using Tensorflow and Keras \cite{KERASblog, VAEblog}. It is to be noted that the output is sensitive to the number of hidden layers and other hyperparameters such as epochs and batch sizes.

\section{Results}
\subsection{Ising Model}
For the Ising model we know that the information about the different phases can be obtained through the order parameter magnetization, which is a scalar. In this case we therefore choose a latent layer with one parameter. 

The input are the spin configurations in a binary form such that spin-up is taken as 1 and spin-down as 0. We pick the sigmoid function as activation for the output layer to minimize reconstruction loss. The activation for the latent layer is the linear function to reduce restrictions on the latent parameter.

We compute the reconstruction loss using the binary cross entropy loss instead of mean-square-error as the input data for Ising model is in binary form. As mentioned earlier, the encoder outputs both the mean, $\mu$, and variance $\sigma$. Here the latent parameter that we study is the mean ($\mu$).

\begin{figure}
\centering
    \includegraphics[width = 8cm]{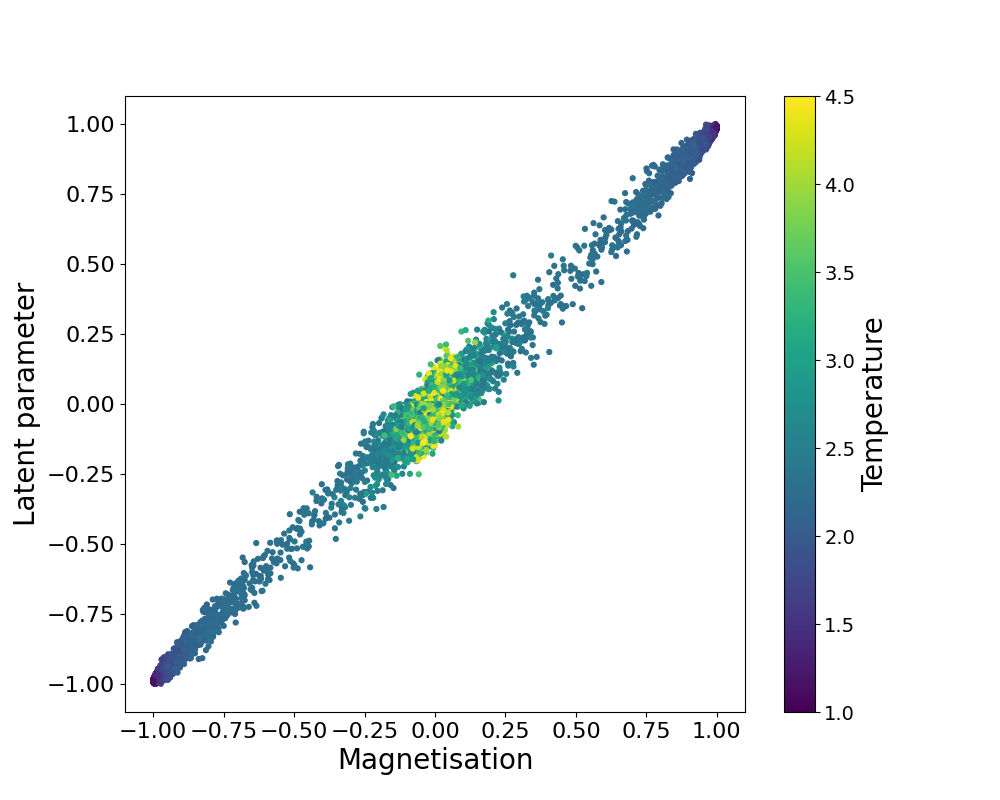}
    \includegraphics[width = 8cm]{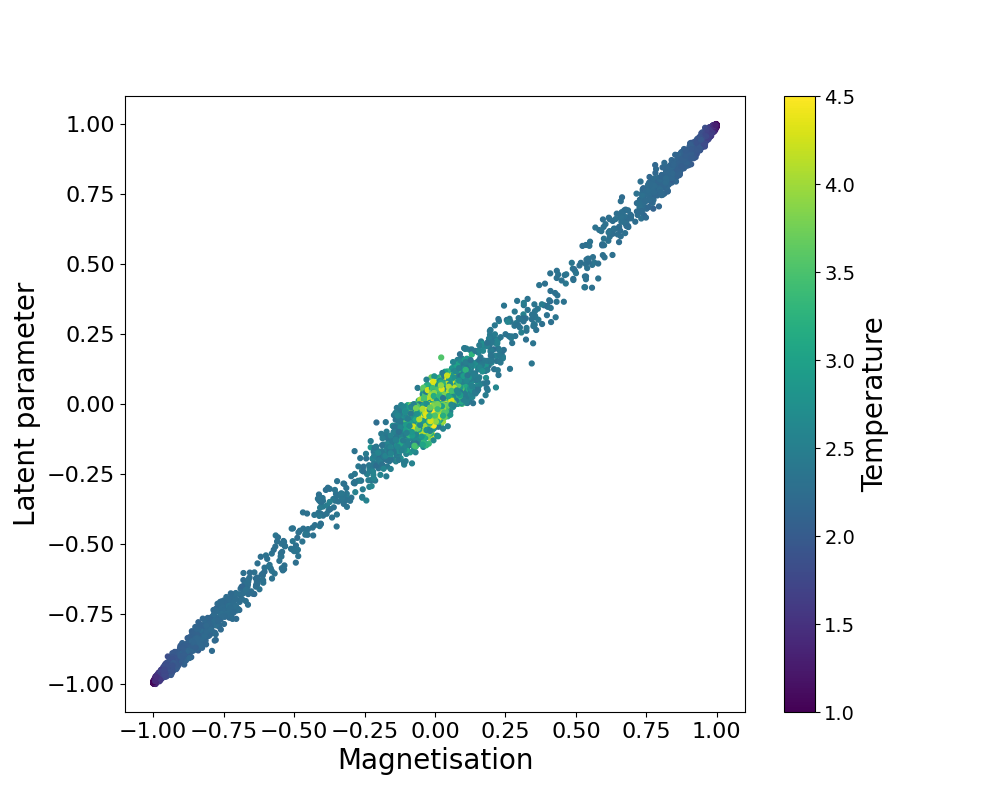}
    \caption{The extracted latent parameter is shown as a function of magnetization for lattice sizes of $L = 50$ (top) and $L = 70$, respectively using a Conditional neural network. The color coding of temperatures is shown by the vertical panes on the right. A linear correlation is clearly visible.}
    \label{fig:latentvsmag}
\end{figure}

In Fig. \ref{fig:latentvsmag}, we show the 
latent parameter with magnetization for lattice sizes with $L = 50$ (top pane) and $L =  70$ (bottom pane). One can observe a clear linear correlation between them, this improves upon the result of a previous work which used non-conditional VAEs \cite{wetzel2017unsupervised}.
Temperature ranges between 1 $-$ 4.5 is shown by the color coding in the vertical panes on the right. 

\begin{figure}
    \centering
    \includegraphics[width = 8cm]{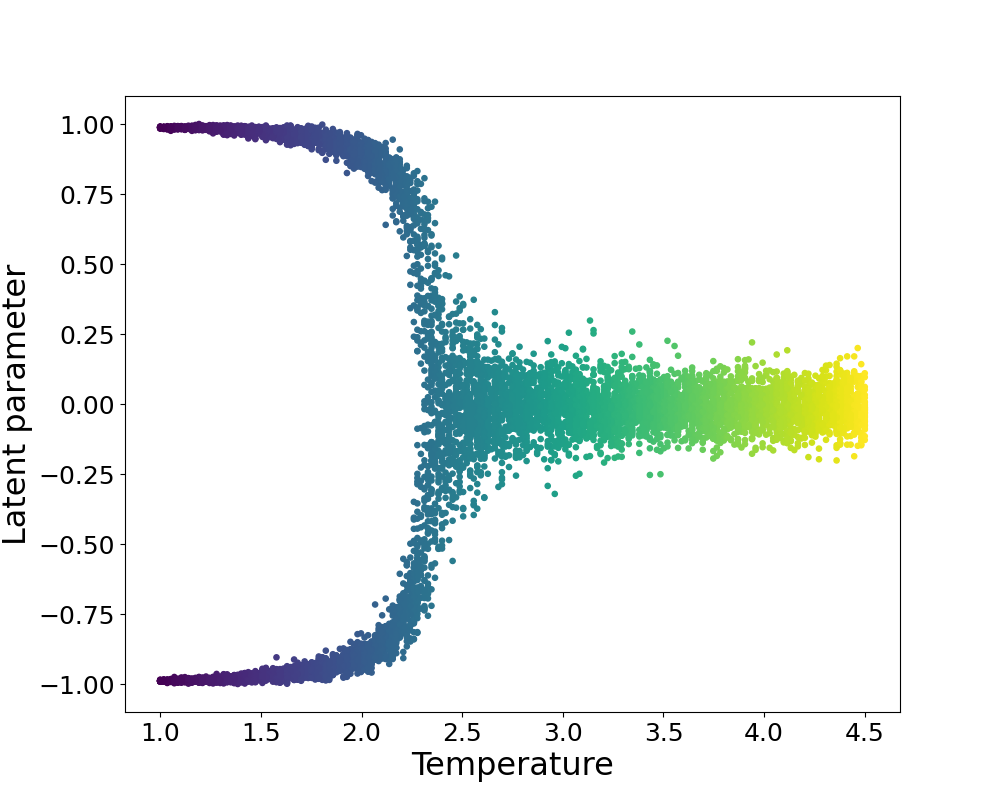}
    \includegraphics[width = 8cm]{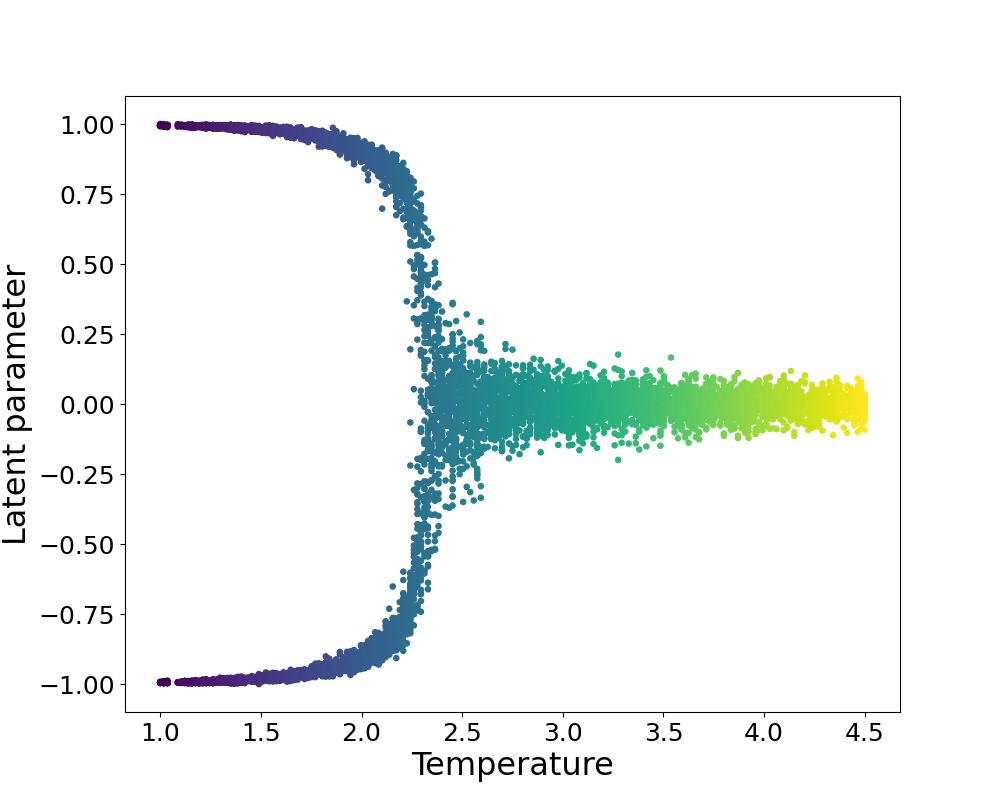}
    \caption{The latent variable extracted from the validation dataset is shown at different temperatures at two different lattice sizes of $L = 50$ (top) and 70 (bottom). Separation of the two phases at lower temperatures is clearly evident.}
    \label{fig:latentvstemp}
\end{figure}

In Fig.  \ref{fig:latentvstemp}, we show the
latent parameter at different temperatures for the validation dataset at two different lattice sizes, $L = 50$ and $L = 70$.
It is evident that the latent parameter is able to separate the two phases effectively at lower temperatures. 
As expected, we also observe that the restoration of the $\mathbb{Z}_2$ symmetry of the Ising model becomes more precise at the larger lattices.


\begin{figure}
    \centering
    \includegraphics[width = 9cm]{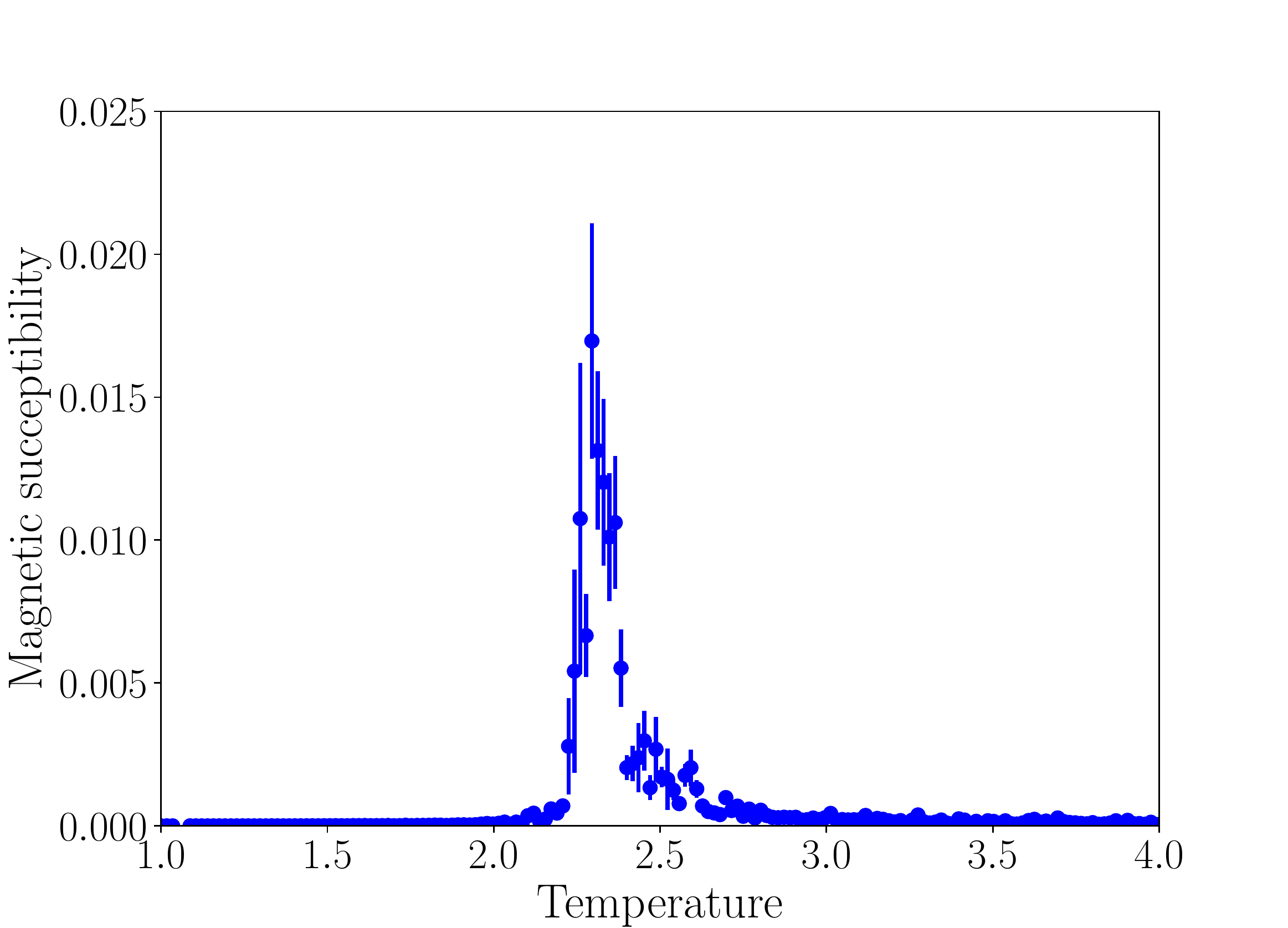}
    \includegraphics[width = 9cm]{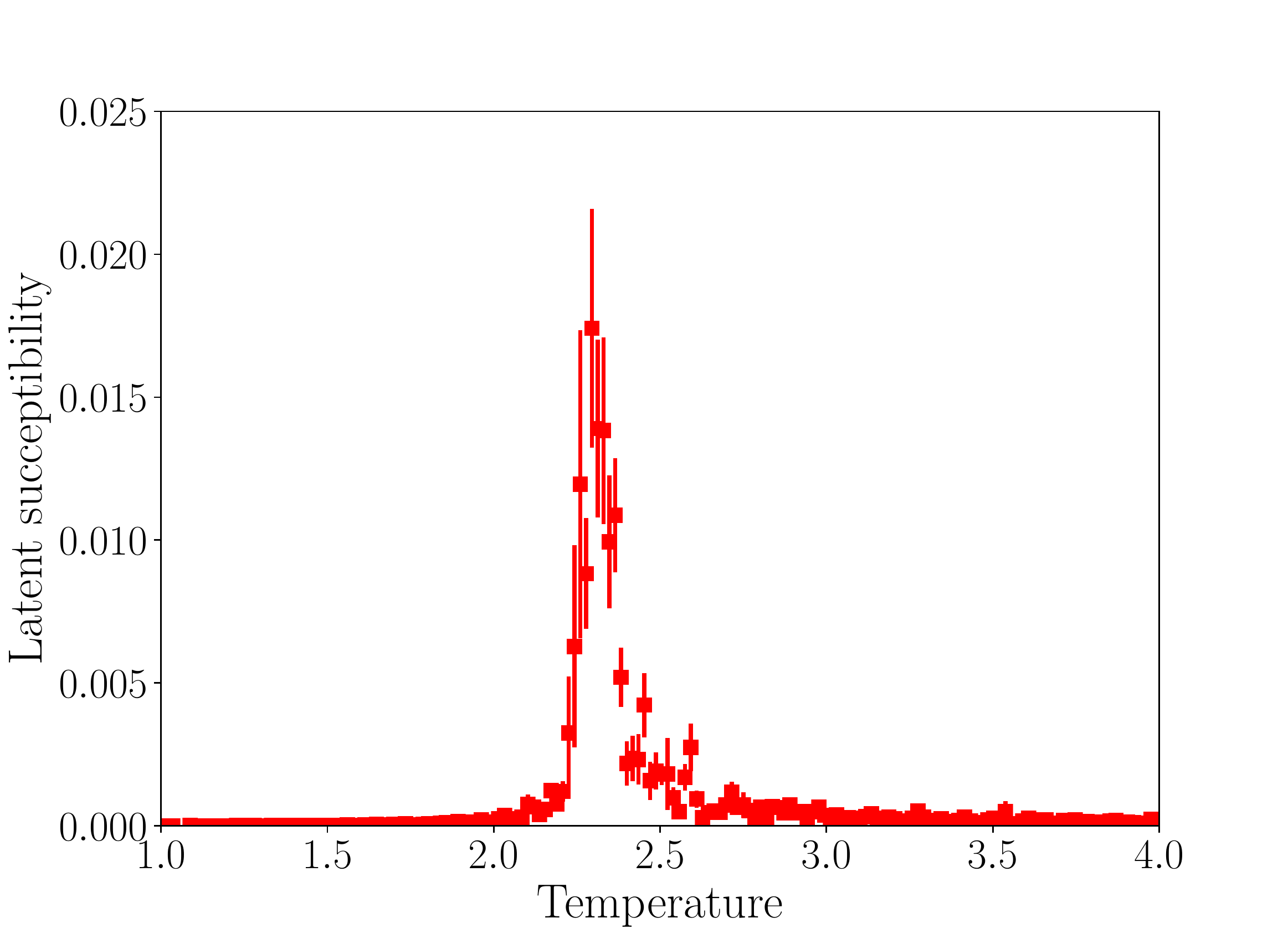}
    \caption{The magnetic susceptibility (top) and the latent susceptibility (bottom) as a function of temperature for L = 70.}
    \label{fig:susceptibility1}
\end{figure}

As usual, the critical point for the Ising model can be extracted using the magnetic susceptibility. 
For the magnetization $m$, the magnetic susceptibility is defined as,
\begin{equation}
    \chi = \frac{L^2}{T} (\langle m^2 \rangle - \langle m \rangle^2 ).
\end{equation}
Following that, here we also define the latent susceptibility corresponding to the latent parameter $z$ as,
\begin{equation}
\chi_z = \frac{L^2}{T}(\langle z^2 \rangle - \langle z \rangle^2).
\end{equation}
In Fig. \ref{fig:susceptibility1}, we show the extracted magnetic susceptibility (top 
 figure: blue circles) and latent susceptibility (bottom figure: red squares) on lattice size $L = 70$. Errorbars are calculated by Jackknifing the procedure. Both of them clearly show the susceptibility peak at the critical temperature. To compare these susceptibilities we plot them together in Fig. \ref{fig:susceptibility1}. It is evident that the above-defined latent susceptibility clearly overlaps with the magnetic susceptibility and their peaks are also found to be at the same temperature range on this bigger-size lattice.  

Next, we calculate the critical temperatures from the respective peak positions of the magnetic and latent susceptibilities at different lattice sizes.
In Fig. \ref{fig:crittempising} we show  the critical temperatures extracted from the magnetic and latent susceptibilities by the blue circles and red squares, respectively. We observe that these susceptibilities approach the correct value of $T_c(L=\infty) = 2.269$ \cite{onsager1944} with bigger lattice sizes.

The results discussed above, and as shown in Figs. 1 - 5, demonstrate that for the case of 2D Ising model the variational autoencoder i) can identify its phases, and ii) can estimate the correct critical temperature. In summary, the latent parameter can replicate the order parameter accurately in this case.

\begin{figure}
    \centering
    \includegraphics[width = 9cm]{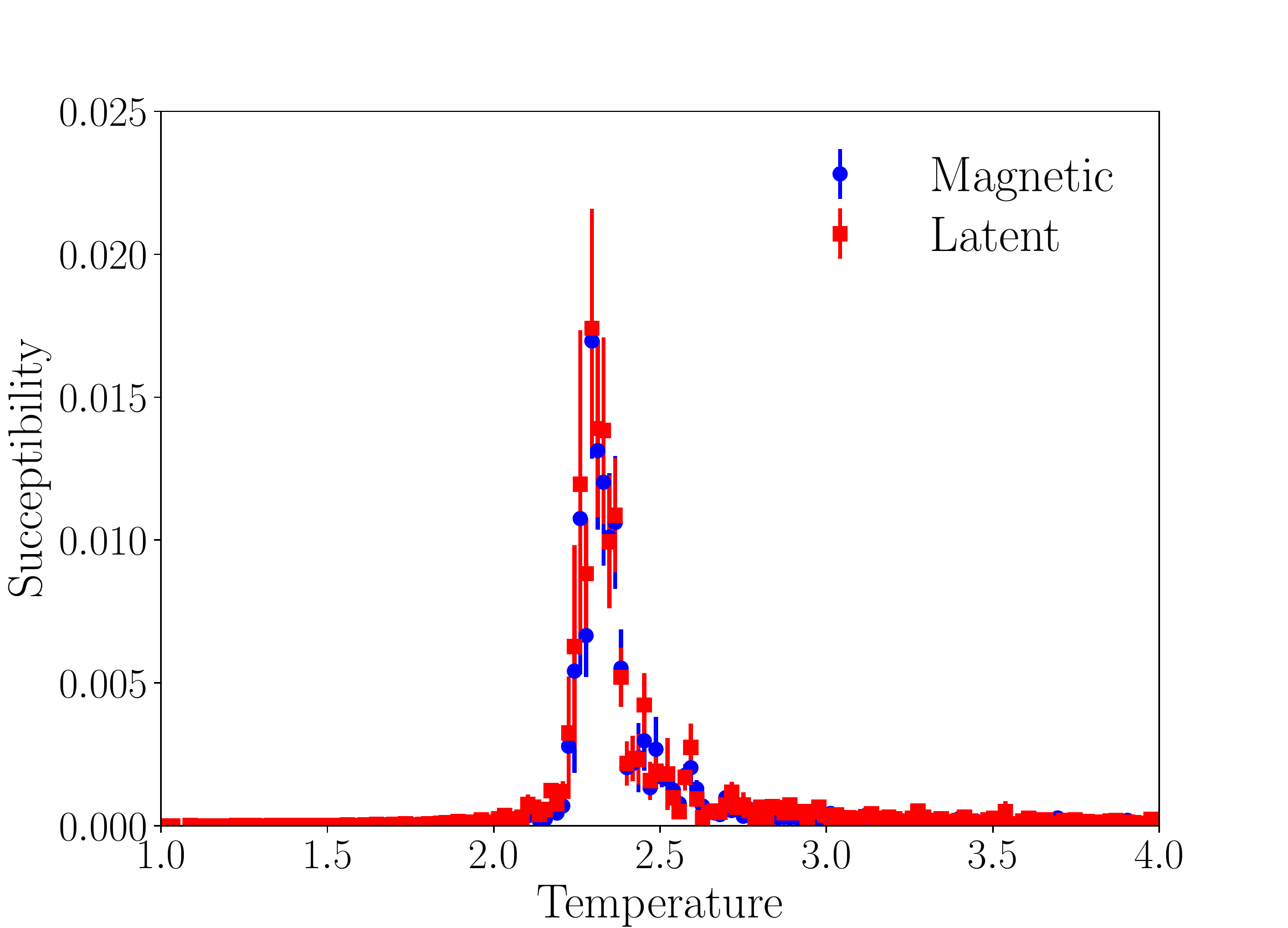}
    \caption{The magnetic susceptibility (top) and the latent susceptibility (bottom) as a function of temperature for L = 70.}
    \label{fig:susceptibility2}
\end{figure}

\begin{figure}
    \centering
    \includegraphics[width = 9cm]{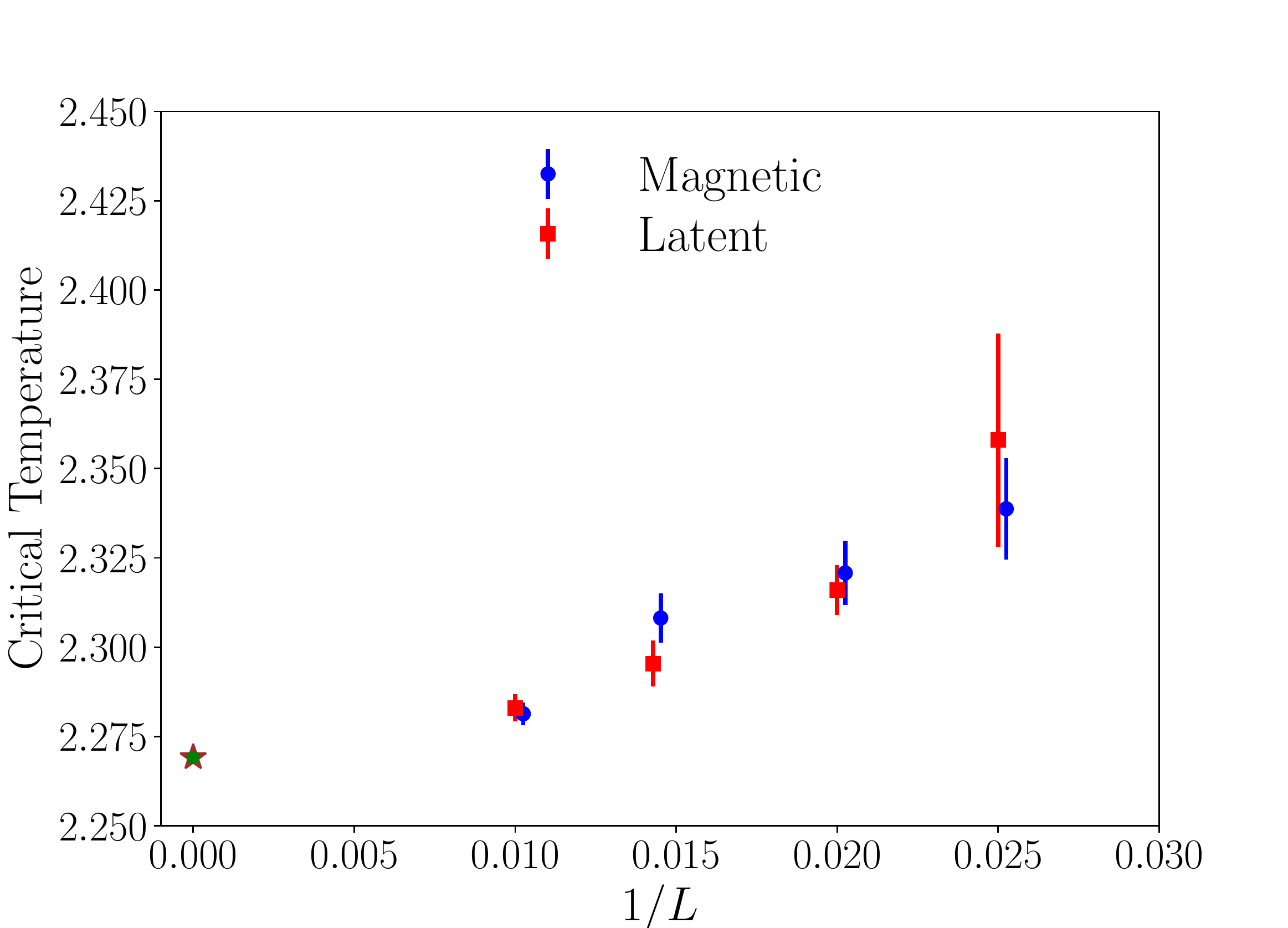}
    \caption{The Critical temperature $T_c(L)$ extracted from the peak of Magnetic and Latent susceptibility. Error bars were obtained from jackknifing the procedure.}
    \label{fig:crittempising}
\end{figure}

\subsection{XY model}
 In this work we have investigated the usefulness of C-VAEs to study the 2D XY model similar to our study of Ising model as discussed in the previous section. However, unlike the Ising model, which displays a discrete $\mathbb{Z}_2$ symmetry, the XY model displays a continuous $O(2)$ symmetry. Hence our task here is to examine the property of the Conditional variational autoencoders to encode continuous symmetries. To be mentioned here that the 3D-XY model exhibits a standard ferromagnetic phase transition from an ordered state at low temperature to an unordered state at high temperature owing to the spontaneous breaking of the $O(2)$ symmetry. Past studies \cite{wetzel2017unsupervised} have analyzed the ability of autoencoders to realize these phases into a latent representation.
In the 2D XY model however, there is a different kind of phase transition,
referred as the Berezinskii-Kosterlitz-Thouless (BKT) transition \cite{kosterlitz1973ordering}, where the system  moves from a disordered high-temperature state to a quasi-ordered state below the critical temperature. We investigate whether VAEs are suitable in identifying these states and the critical temperature. 

In this case the input layer consists of $2\times L^2$ neurons, $L$ being the lattice size, thus including both the X and the Y components of individual spins. We use a latent layer with two components as it gives the appropriate representation.
 In this case an order parameter can be defined as the norm of a two-component vector. We define the L2 squared norm of the latent parameter ($||z||^2 = z_x^2 + z_y^2$) vector which can correspond to the squared magnetization ($||m^2|| = (\sum s_x)^2 + (\sum s_y)^2$).

\begin{figure}
    \centering
    \includegraphics[width = 9cm]{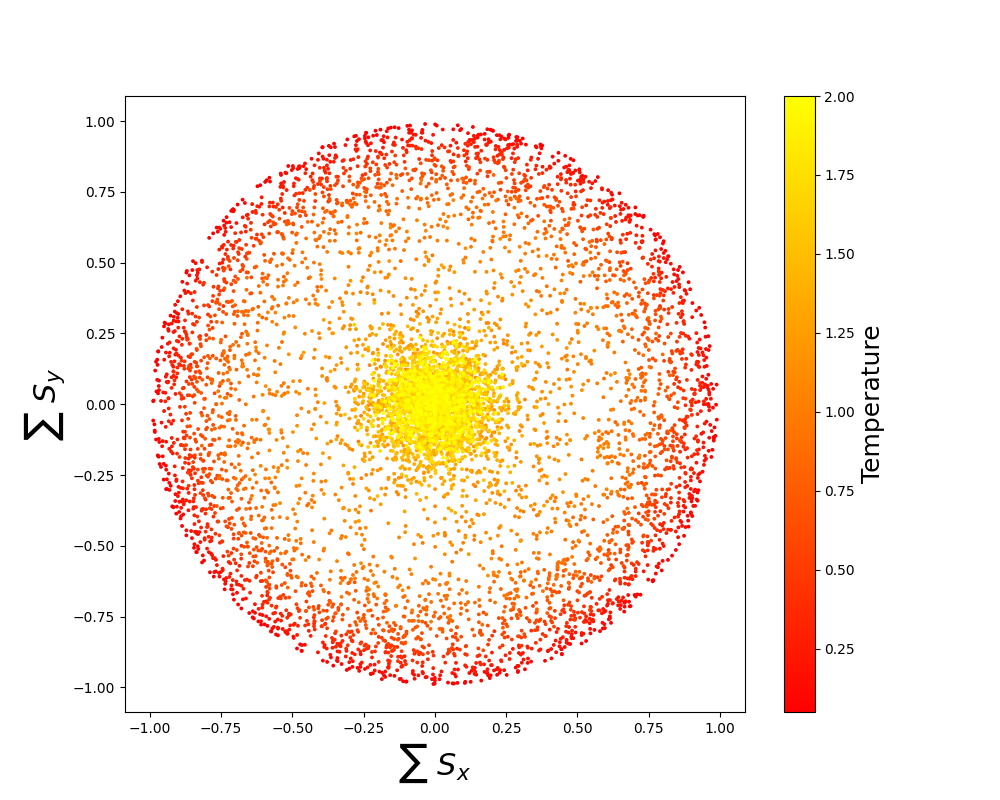}
    \includegraphics[width = 9cm]{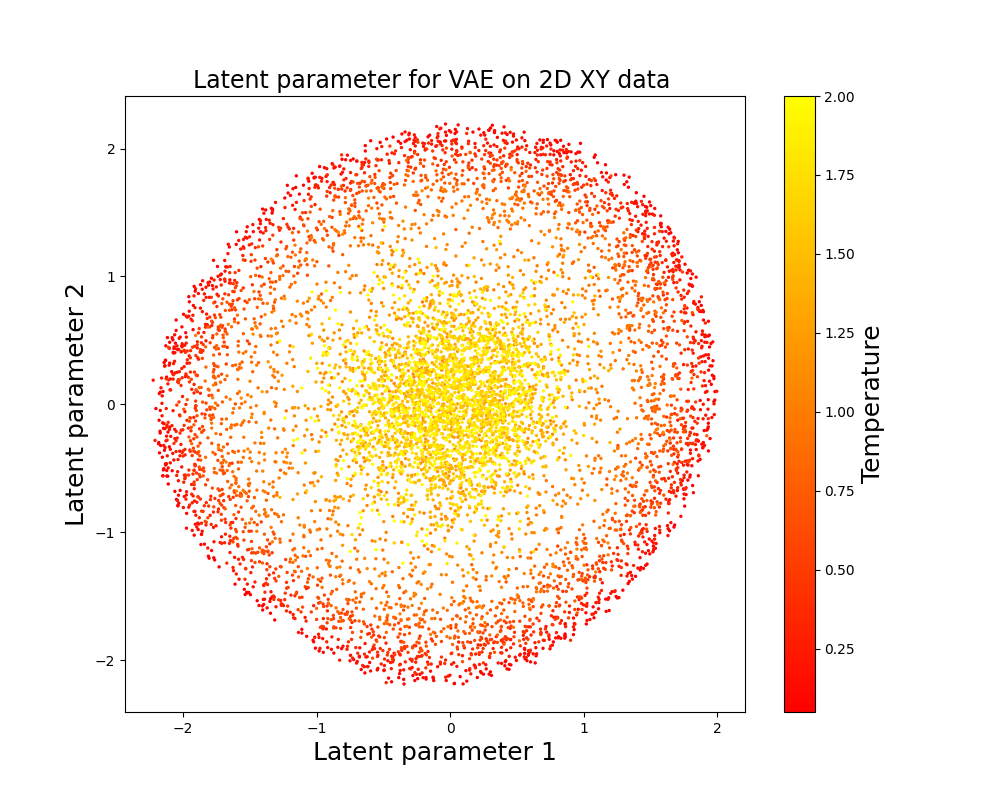}
    \caption{(Top) The X-component against the Y-component of the total spin of each configuration and (Bottom) the two dimensional latent space.}
    \label{fig:xymagxy}
\end{figure}

\begin{figure}
    \centering
    \includegraphics[width = 7cm]{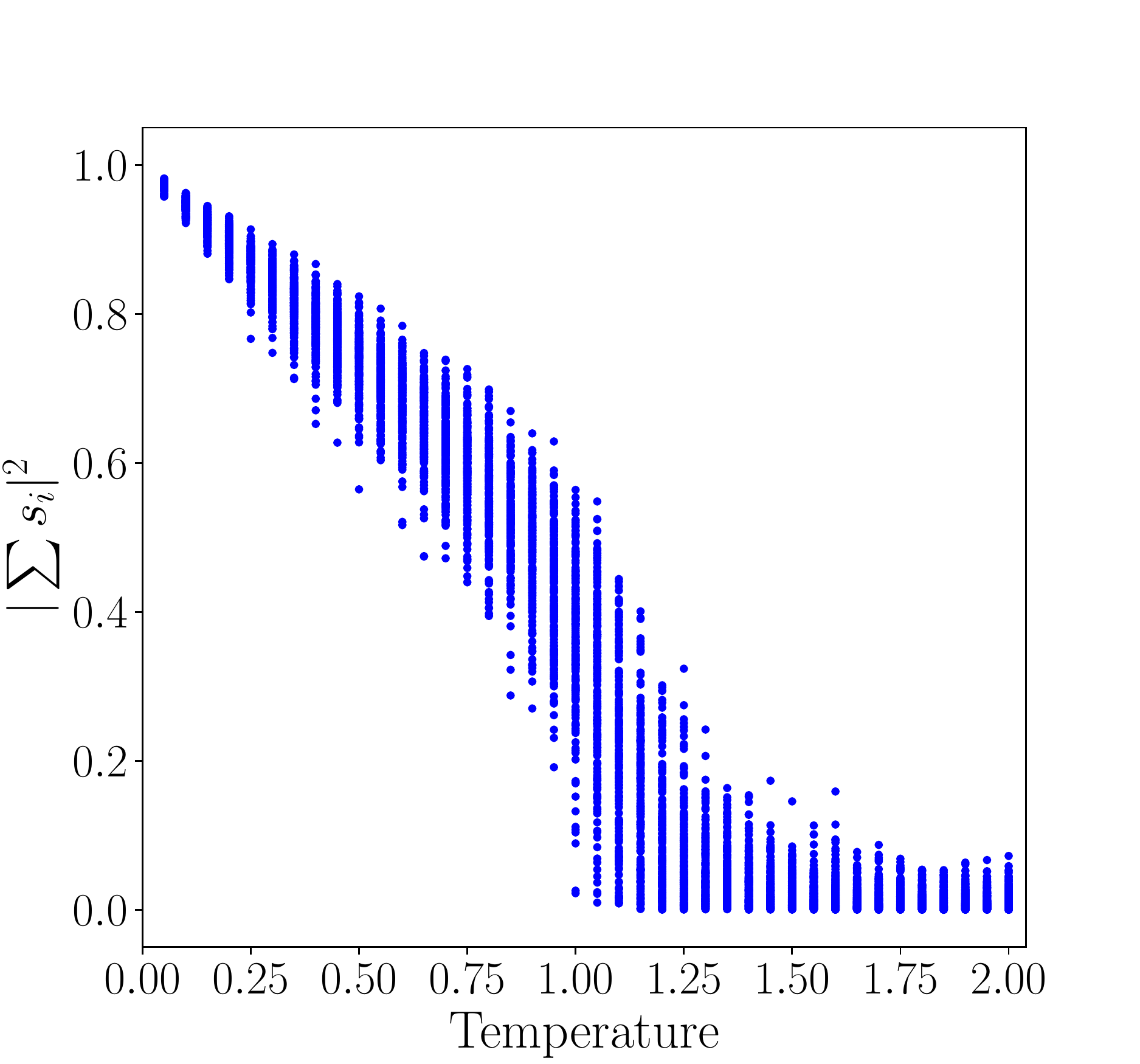}
    \includegraphics[width = 7cm]{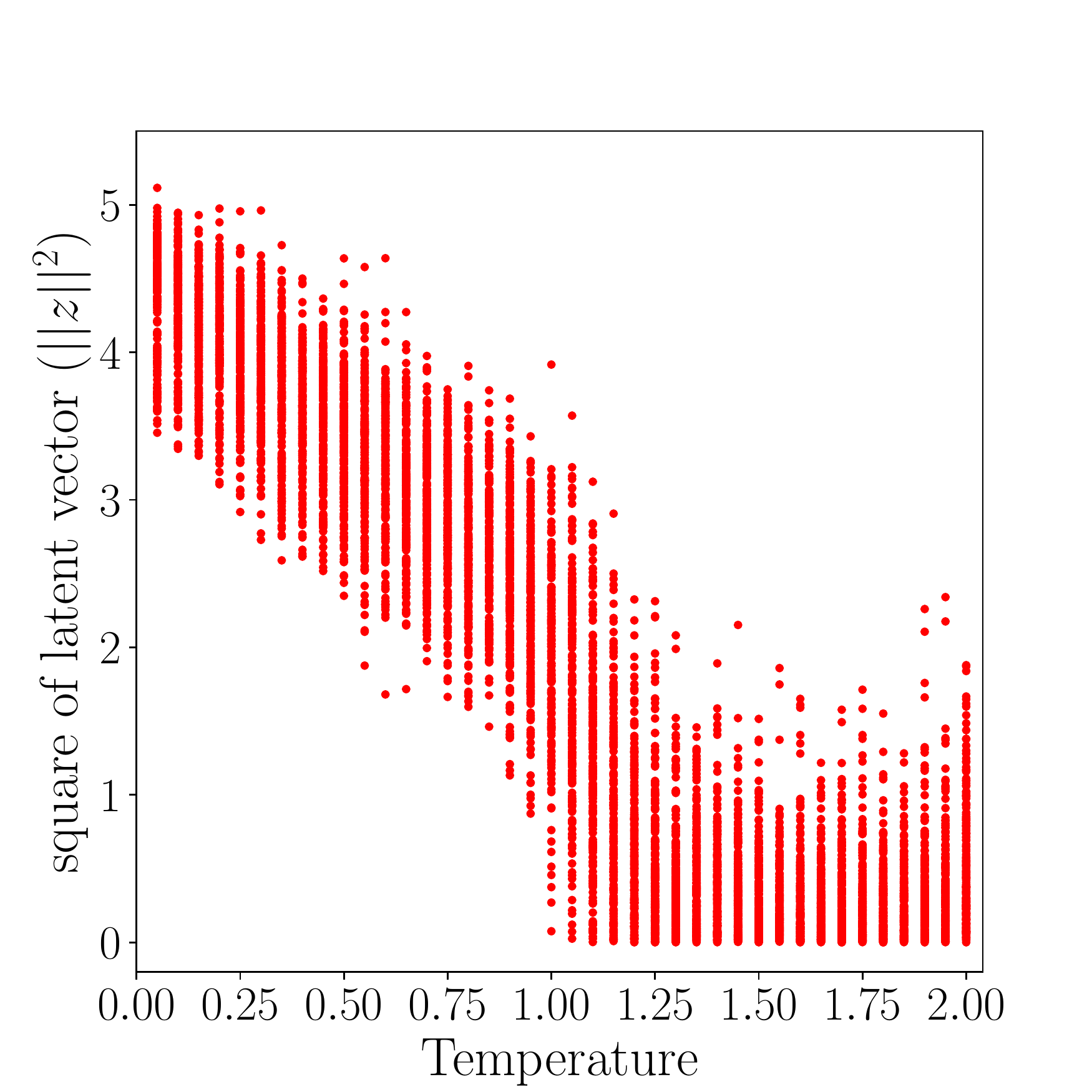}
    \caption{(Top) Squared magnetization vs. Temperature and (Bottom) L2 Norm squared of the latent vector for the testing dataset.}
    \label{fig:xyorderlat}
\end{figure}

In Fig \ref{fig:xymagxy}, we show the  distribution of the sum of spin components as well as the latent variables. Similar to the components of spin-sum ($\sum S_i$), the latent parameters in this case also clearly shows the separation of the high temperature disordered phase from the low temperature quasi-ordered phase. 
To make a correspondence between the magnetization and the latent parameters, in Fig. \ref{fig:xyorderlat} we show  
$||m^2||$
and $||z||^2$. It is interesting to find that for temperatures below $T = 2$ their variation with temperatures are quite similar. However, for higher temperatures they are observed to be different.

\begin{figure}
    \centering
    \includegraphics[width = 9cm]{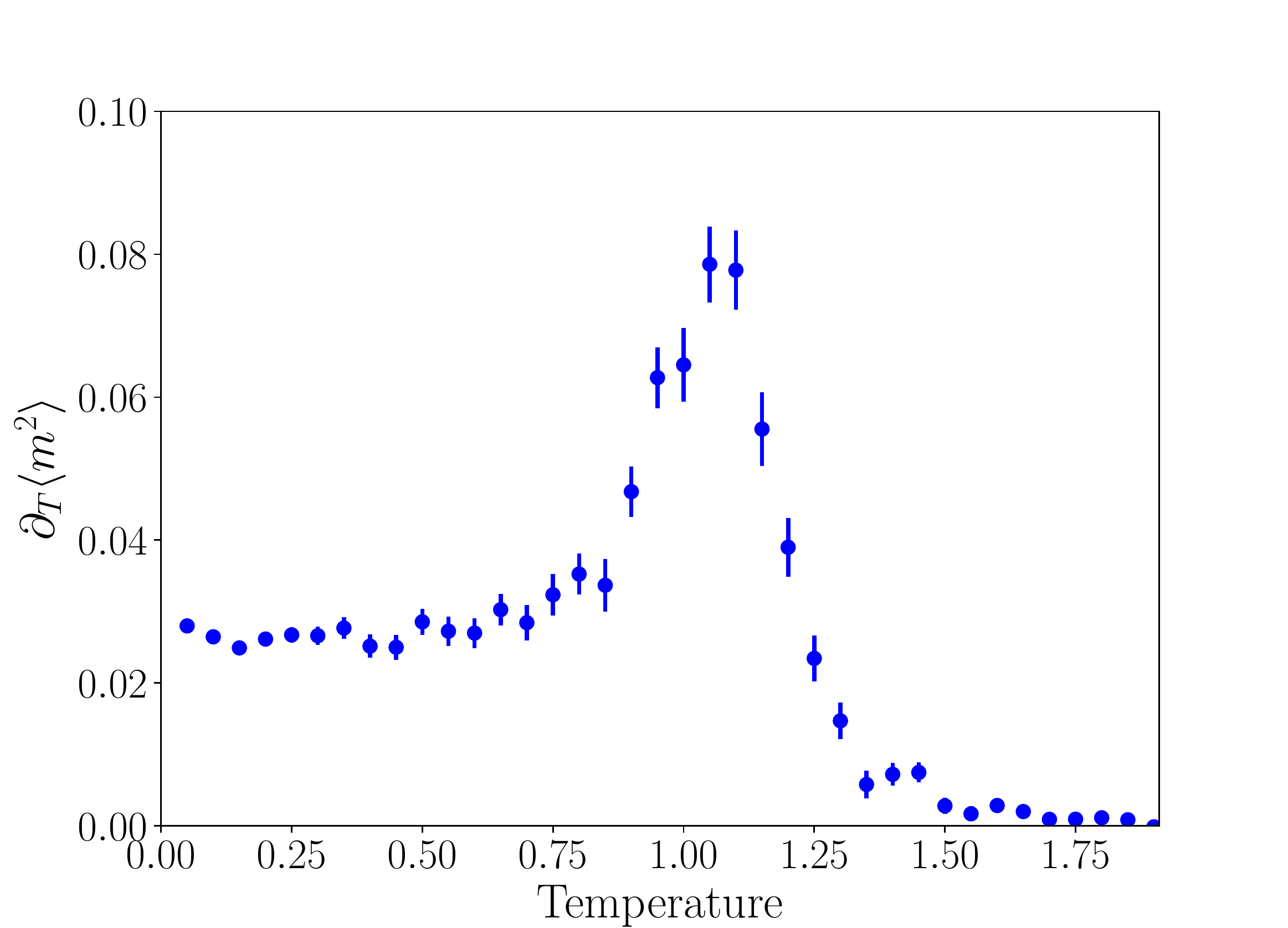}
    \includegraphics[width = 9cm]{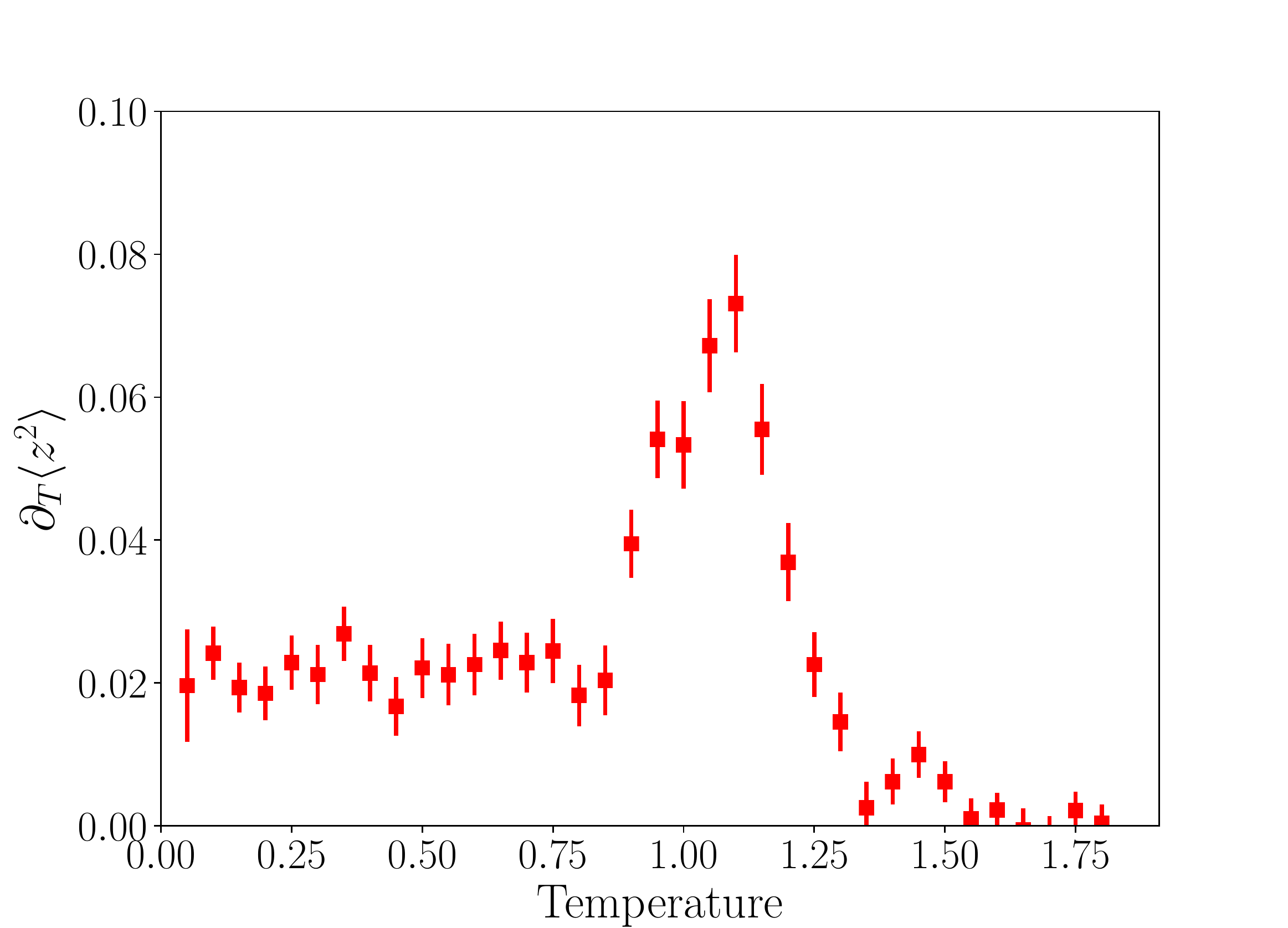}
    \caption{The derivative of the average norm squared of Magnetization vector (top) and latent vector (bottom) with respect to temperature.  }
    \label{fig:xysus1}
\end{figure}

\begin{figure}
    \centering
    \includegraphics[width = 9cm]{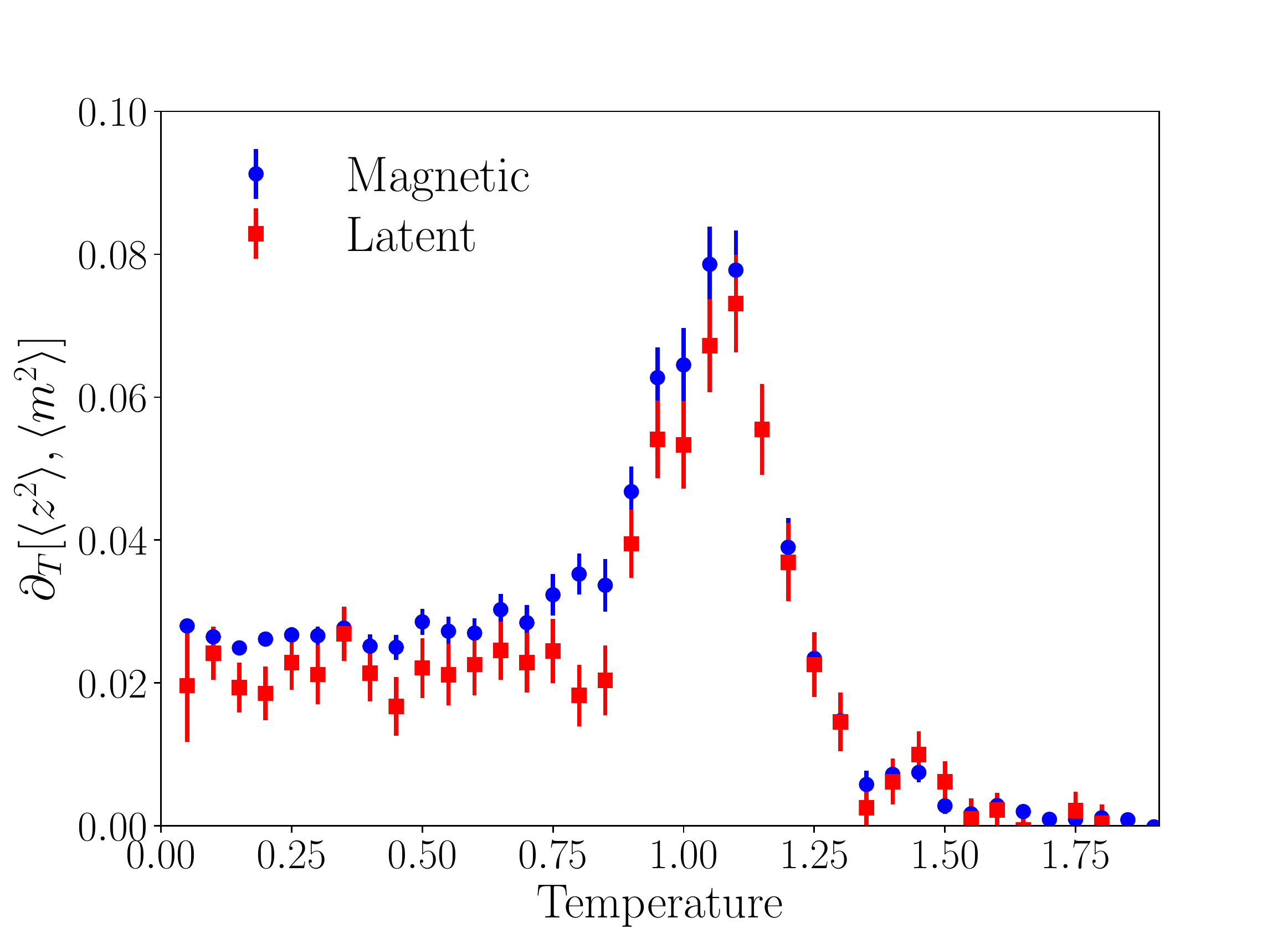}
    \caption{The derivative of the average norm squared of Magnetization vector (top) and latent vector (bottom) with respect to temperature.  }
    \label{fig:xysus2}
\end{figure}

Here again we calculate the critical point 
using the peak of the magnetic susceptibility as well as from the latent parameter. For the latent parameter we use the negative gradient of the average L2 norm of the latent vector to get a correspondence with the magnetic susceptibility. In Fig. \ref{fig:xysus1} we show them by the top and bottom plots, respectively, and in Fig. \ref{fig:xysus2} we plot them together showing their correspondence as a function of temperatures. It is clear that the two magnetic and latent susceptibilities follows each other till a large temperatures and their peaks are also more or less at the same temperatures. 
We obtain a critical temperature of $T_{c,\ latent} = 0.960 \pm 0.03$ from the latent parameter as compared to $T_{c, \ mag} = 1.063 \pm 0.0065$ from the magnetic susceptibility, which are close to the theoretical prediction of $T_c = 0.8816$.
In summary, above results demonstrate that C-VAEs can be utilized effectively for identifying different phases in 
$2D$ XY model and a correspondence can be made between the magnetic and latent susceptibilities.

\section{Conclusion}
In this work, we have employed Conditional Variational Autoencoders
to study two-dimensional Ising as well as two-dimensional XY models. Spin configurations are
 generated using Wolf cluster algorithms. Our main objective was to investigate the usefulness of C-VAEs in distinguishing different phases of these models as a function of temperatures.  We have shown that the utilization of C-VAEs resulted in a notable improvement in accuracy compared to traditional AEs and VAEs trained on configurations across all temperatures for the restoration of the $\mathbb{Z}_2$ symmetry \cite{wetzel2017unsupervised}. Furthermore, the latent parameter has exhibited the ability to approximate the relevant order parameters with reasonable accuracy. An interesting outcome of our work is the successful determination of the critical temperatures using the trained networks. For the Ising model, we observe that as the lattice size increases, the critical temperature estimated from the latent parameter converges towards its correct value. In the case of the XY model, the critical temperature obtained from the latent parameter is found to be closer to its theoretical value compared to the estimation derived from Monte Carlo simulations. 
 
Considering the symmetries inherent in these models, we make specific choices for the dimensionality of the latent parameter. For the Ising model, a single latent parameter is chosen, while for the XY model, a two-dimensional latent parameter is employed. Following this  a three-component latent vector could potentially be utilized to study the O(3) or SU(2) gauge theory \cite{wetzel2017machine}.

Overall, our work highlights the effectiveness of Conditional Variational Autoencoders in characterizing different phases for the 2D Ising and XY models. Our findings provide valuable insights about the critical temperatures of the selected models and pave the way for possible further investigations into the application of C-VAEs in other models relevant to condensed matter physics.


\section*{Acknowledgments}
This work is supported by the 
Inspire fellowship of Department of Science and Technology, and the 
Department of Atomic Energy, Government of India, under Project Identification Number RTI 4002.  We thank D. Banerjee, S. Paul, P. Srivastava  and Utkarsh for discussions. Computations were performed at the GPU cluster of the Department of Theoretical Physics, TIFR.


\bibliography{cvae}

\section*{Appendix}
Here we provide a general discussion on autoencoders, variational autoencoders and conditional variational autoencoders following the literature of AI. Autoencoders are a variant of traditional feed-forward neural networks utilize for learning data in an unsupervised manner \cite{autoenc}. It consists of an encoder that encodes the input data $({x})$ into a latent variable $({z})$ and a decoder network that decodes the latent variable into an approximation of the input. Essentially an autoencoder consists of two functions, the encoder $f_{\theta}$ and the decoder $g_{\phi}$ such that $x = g_{\phi}(f_{\theta}(x))$. Autoencoders can also be employed as non-linear methods for dimensional reduction. Conditional autoencoders \cite{sohn2015learning} are modified autoencoders such that the input of the decoder is the output of the encoder along with a label for the input data {\it i.e.}, for an input data $x$ with label $y$, the reconstruction of the input is of the form $x = g_{\phi}(f_{\theta}(x), y)$.

\subsection*{Variational Autoencoders}
Below we give a general discussion on variational autoencoders (VAE) following Refs. \cite{doersch2016tutorial, kingma2019introduction}. VAEs are a version of autoencoders that impose additional constraints on the encoded representation. Unlike autoencoders, which learn an arbitrary function to encode and decode data, VAEs learn the parameters of a probability distribution modeling the data. This helps to generate samples closely resembling the input data once the network is trained. Essentially, the functions $f_{\theta}$ and $g_{\phi}$ are taken as probability distributions that the network learns.

VAEs are based on the assumption that the input data can be sampled from a unit Gaussian distribution of latent parameters {\it i.e.}, the unknown data distribution $\hat{p}$ can be described by random vectors $z \in \mathbb{R}^d$ and a prior $p(z)$ in a low dimensional latent space. Here $d$ is the dimension of the latent variable.

The encoder network maps the input to the parameter $(\mu(x), \sigma(x))$ of the latent Gaussian representation $f_{\theta}(z|x)$ of $p(z|x)$, where $\mu(x)$ and $\sigma(x)$ represent the mean and standard deviation vectors of the model. The decoder network uses the samples of the Gaussian  $f_{\theta}(z|x)$ and a label of the input data as input to generate new samples according to the distribution $g_{\phi}(x|z)$. All parameters are then estimated by maximizing the bound of the marginal log- likelihood $\log p_{\theta}(x)$. The input of the decoder $(z)$ is defined using the re-parameterization trick such that \cite{doersch2016tutorial, kingma2019introduction} 
\begin{equation}
    z = \mu + \sigma \otimes \mathcal{N},(0,1)
\end{equation}

In this work, the latent parameter that we employ to model physical systems is the mean $\mu(x)$. In the training phase, the network learns the parameters $\theta$ and $\phi$ such that $x \sim g_{\phi}(f_{\theta}(x))$. To monitor the learning progress of the VAE, we keep track of the \textit{reconstruction error}. The most simple and commonly used metric for error is the mean-squared error (the squared Euclidean distance between the input and the output data). For $n$ data points the mean-squared-error is \cite{doersch2016tutorial, kingma2019introduction},
\begin{equation}
    E(\theta, \phi) = \frac{1}{n} \sum_{i=1}^{n}(x_i - g_{\phi}(f_{\theta}(x_i)) )^2 .
\end{equation}

Another way to monitor the loss is by using the \textit{binary cross entropy} (BCE), defined as \cite{doersch2016tutorial, kingma2019introduction},
\begin{equation}
    R(x) = - \sum_i^n x_i \log p_i(x) + (1-x_i) \log(1 - p_i(x)),
\end{equation}
where $x_i \in \{ 0,1 \}$ and $p_i(x)$ is the reconstruction probability of the $i^{th}$ bit. 
This definition of loss is especially good for binary input data.

Along with the reconstruction loss, the set of parameters $\theta$ also needs to minimize the difference between the machine distribution $p_{\theta}$ and the data distribution $\hat{p}$. The dissimilarity between the two distributions can be quantified using the \textit{Kullback-Liebler} (KL) divergence \cite{KL-d}. In most unsupervised learning problems, one does not know the distribution underlying a given dataset. From the given samples  it is not possible the determine the KL divergence directly. Instead, we use the nearest neighbor estimation of KL loss. Given two continuous distributions $p$ and $q$ defined on the same probability space, the KL divergence is determined by \cite{doersch2016tutorial, kingma2019introduction},
\begin{equation}
    D_{KL}(p || q) = \sum_{x \sim p(x)} p(x) \log \frac{p(x)}{q(x)}.
\end{equation}
For a conditional network, the input of the decoder is simply chosen as $(z,y)$ where $y$ is the label associated with the input image as discussed before. The process of the VAE is modified a bit such that given the label $y$, $z$ is drawn from the prior distribution $p(z|y)$, and the output $x$ is generated from the distribution $p(x|y,z)$. For a simple VAE, the prior is $p(z)$ and the output $x$ is generated from $p(x|z)$.

In summary, the main idea behind VAEs is to learn a probability distribution over the latent space that can generate data samples similar to the input data. This is achieved by minimizing the reconstruction error between the original data and the reconstructed data, while also minimizing the KL divergence between the learned latent distribution and a prior distribution. The latter term encourages the learned latent distribution to be similar to the prior distribution, which can be a simple Gaussian distribution.

\end{document}